\documentclass[12pt,a4paper,oneside]{article}
\usepackage{amsfonts,amssymb,amsmath,amsthm,graphicx}

\title{{Casimir energy
\\and realistic model of dilute dielectric ball}}
\author{Valery N. Marachevsky \thanks{E-mail: root@VM1485.spb.edu}}
\date{{\normalsize Department of Theoretical Physics\\St.Petersburg State
University \\198904 St.Petersburg , Russia}\\  }

\begin{document}
\newcommand{\om}{\bigl|\omega\bigr|\sqrt{\varepsilon}}
\newcommand{\omm}{\bigl|\omega\bigr|}

\maketitle

\begin{abstract}

The Casimir energy of a dilute homogeneous nonmagnetic dielectric
ball at zero temperature is derived analytically for the first
time for an arbitrary physically possible frequency dispersion of
dielectric permittivity $\varepsilon(i\omega)$. A microscopic
model of dielectrics is considered, divergences are absent in
calculations because an average interatomic distance $\lambda$ is
a {\it physical} cut-off in the theory. This fact has been
overlooked before, which led to divergences in various macroscopic
approaches to the Casimir energy of connected dielectrics.

\end{abstract}

\begin{flushleft}
keywords: dipole interaction, Casimir energy, dilute dielectrics
\end{flushleft}

\newpage

\section{Introduction}

In a recent paper \cite{Mar} we proposed a technique for the
calculation of Casimir energies of connected dilute dielectric
bodies based on a microscopic structure of dielectrics. An
interatomic distance  (denoted here by $\lambda$) is a
physical cut-off within this approach, the Casimir energy
is considered in the second order $(\varepsilon(i\omega) -
1)^2$ - the lowest order for the energy of interaction of atoms
constituting the ball. The equivalence of a Casimir energy and
a dipole-dipole interaction for homogeneous dielectrics of an arbitrary
form in the order $(\varepsilon(i\omega) - 1)^2$ was proved in \cite{Mar} on
the basis of Lifshitz theory \cite{Lif1}\cite{Lifshitz}, so the
energy of a dielectric due to a dipole-dipole pairwise
interaction of atoms constituting the ball coincides in this order
with the Casimir energy (see a line of reasoning in section $2$ of the
present paper).

Macroscopic and microscopic approaches  to the Casimir energy with
emphasis on the case of dilute connected dielectrics are discussed
in section $2$. In section $3$ we use the technique suggested in
\cite{Mar} to derive  for the first time the formula for the
Casimir energy of a homogeneous dilute dielectric ball in the
order $(\varepsilon(i\omega) - 1)^2$, valid at zero temperature
for an arbitrary physically possible frequency dispersion of the
ball dielectric permittivity $\varepsilon (i\omega)$. It is proved
that Casimir surface force on a dilute dielectric ball is
attractive for any $\varepsilon(i\omega)$. In the hypothetical
non-physical limit $\lambda \to 0$ the Casimir energy is divergent
for every model of $\varepsilon(i\omega)$. This explains the fact
that without the interatomic distance $\lambda$ every approach to
the Casimir energy of connected dielectrics was ill-defined. In
fact, in condensed matter physics the interatomic distance often
serves as a physical cut-off, which makes expressions finite and
well-defined. It is argued in this paper that from this viewpoint
the Casimir effect is not different from other areas of
condensed matter physics, the interatomic distance $\lambda$ is an
important ingredient of the theory of connected dielectrics,
it makes the theory finite and all calculations are well-defined.

The Casimir energy of a dilute dielectric
ball is still believed to be equal to
\begin{equation}
E_1 = \frac{23}{1536\pi a} (\varepsilon - 1)^2 , \label{ad1}
\end{equation}
which is obviously only a non-dispersive limit of one of the terms
in the full expression for the energy (\ref{tt3})
(see Eq.(\ref{tt5}), where the term (\ref{ad1}) is derived from
the fourth line of (\ref{tt3})). The full energy expression
(\ref{tt3}) contains additional terms: volume
and surface contributions to the energy, which are the first and second
lines of Eq.(\ref{tt3}) respectively, as well as the terms on the
third line of Eq.(\ref{tt3}), which
don't depend on the ball radius $a$.
Below in the introduction it is our aim to explain why the term (\ref{ad1})
is only the large distance contribution to the Casimir energy of the
ball, discuss the subtle points of some approaches where the
term (\ref{ad1}) was derived and try to persuade the reader why
the  correct expression for the Casimir energy of a
dilute dielectric ball is given by Eq.(\ref{tt3}).

Only when the energy is defined as in (\ref{tt2}) and (\ref{tt3}), it is equal
to the sum of dipole-dipole pairwise interactions between the
atoms of the ball. The Casimir-Polder potential
\begin{equation}
 U_{Cas.-Pol.}= -\frac{23 \alpha_1(0)\alpha_2(0)}{4\pi r^7} ,
 \label{ad2}
\end{equation}
which is a large distance limit of a dipole-dipole potential
$(\ref{tt1})$, valid only for distances $r\gg 1/\omega_0$,
$\omega_0$ is a characteristic absorption frequency of atoms, was
used instead of a dipole-dipole potential (\ref{tt1}) in the
calculation \cite{M1}, where the term (\ref{ad1}) was first
derived. The calculation \cite{M1} correctly extracted the large
distance contribution to the energy with a dispersion neglected.
However, the calculation \cite{M1} was non-dispersive from the
outset and thus couldn't yield the contribution to the energy from
short distances ($r \lesssim 1/\omega_0$) between atoms of the
ball. Moreover, the use of a dimensional regularization in
\cite{M1} concealed the divergences which would appear in the
energy expression from the integration over short distances
between atoms in the 3-dimensional ball since the minimum distance
between atoms of the ball was not introduced in \cite{M1}.

It is often said that
the result (\ref{ad1}) is the unique finite Casimir energy.
In our opininon, this statement demonstrates the failure
to appreciate the physically reasonable definition of the Casimir
energy. This should be clear from the line of
the following examples.
The Casimir energy of two neutral atoms coincides with the energy of a
dipole-dipole interaction of these atoms. When an atom is located
outside the dielectric of an arbitrary form,
then in the dilute approximation the Casimir energy is
equal to the sum of dipole-dipole interactions between the atom
and atoms of the dielectric \cite{Mar}. For two parallel dielectric slabs
it is well known that the Casimir energy in a dilute approximation
is equal to the sum of pairwise dipole-dipole
interactions of atoms constituting the slabs. For a dilute dielectric
ball the sum of pairwise dipole-dipole interactions of atoms
constituting the ball is given by Eq.(\ref{tt3}), not Eq.(\ref{ad1}).

Needless to say that the term (\ref{ad1}) itself was really
important for development of the theory of the Casimir effect in
connected dielectrics since this term has been derived via
different techniques \cite{M1, tech}. These techniques included both
field-theoretic Casimir calculations \cite{tech} and
the summation of the Casimir-Polder potential in $d$ dimensions
\cite{M1} discussed above. However, from the field-theoretic approach applied
as in \cite{tech} one can extract correctly only the large distance
contribution to the Casimir energy of the ball, e.g.$E_1$. This large
distance contribution $E_1$ was found to be the same when summing up the Casimir-Polder
potential (\ref{ad2}) between atoms of the ball \cite{M1} and when the
Casimir energy was derived by field-theoretic calculations \cite{tech} - so the
equivalence of large distance parts of the Casimir energy of a
dilute dielectric ball derived
by microscopic and macroscopic approaches was proved.  It is important
to stress that up to now field-theoretic methods didn't yield
satisfactorily short distance contributions to the Casimir energy
of connected dielectrics. The reason is simple: these methods
were developed for {\it disjoint}, not connected dielectrics,
and application of these methods to connected
dielectrics without any changes inevitably leads
to different types of divergences in {\it every}
field-theoretic calculation of the Casimir energy, these divergences are
reminiscents of the ill-defined short distance structure of the
theory. In this paper we do not aim to present the field-theoretic
macroscopically correct derivation of the ball Casimir energy.
We rather wish to present for the first time the correct {\it microscopic}
derivation of the Casimir energy of a dielectric ball in a dilute
approximation - in the order $(\varepsilon(i\omega) - 1)^2$.

So the object of study is a dispersive dielectric ball of radius
$a$ at zero temperature  which is dilute, i.e. the permittivity
$\varepsilon (i\omega)$ of the ball satisfies the condition
$\varepsilon(i\omega) - 1 \ll 1$ for all frequencies $\omega$.
The Casimir energy is calculated in the second order
$(\varepsilon(i\omega) - 1)^2$. A dipole-dipole potential yields
Casimir-Polder and van der Waals potentials in the large and short
distance limits respectively, we adopt this terminology. We put
$\hbar=c=1$ and use rationalized Gaussian units for the Maxwell
field where polarizability $\alpha(i\omega)$ is defined by
$\varepsilon(i\omega) - 1 = 4\pi \rho\,\alpha(i\omega)$, $\rho$ is
a number density of atoms.

\section{Macroscopic and microscopic approaches to Casimir energy}

The study of Casimir energy for spherical dielectrics began with
the work by Boyer \cite{Boyer} where he derived Casimir energy of
perfectly conducting spherical shell. Numerically it is equal to
$\simeq 0.09235/2a$ \cite{Milton2}. The study of a dielectric ball
and its Casimir energy for an arbitrary permittivity began with
the work by Milton \cite{Milton1} and continues up to now.
Different mathematical methods were used in the Casimir effect
calculations \cite{different}. We are not planning to give a
description of these methods here, we rather wish to emphasize the
main problem in a macroscopic formulation of the theory, discuss
the reason for its appearance, outline the solution for dilute
dielectrics and the area for further research.

For dispersive dielectrics a {\it macroscopic} formulation of the
Casimir effect was originally proposed by Lifshitz \cite{Lif1} and
developed by Dzialoshinskii, Pitaevskii and others
\cite{Lifshitz}. In the macroscopic formulation the system is
usually described by the position and frequency dependent
dielectric permittivity $\varepsilon(\omega,\bf{r})$. For disjoint
dielectrics the theory worked perfectly. The well known case of
dielectric parallel plates became the classic example
\cite{Lifshitz}. However, serious problems were found when this
theory was applied to connected dielectrics. Problems were first
found in the case of a homogeneous nonmagnetic dielectric ball
described by a dielectric permittivity $\varepsilon(\omega)$
\cite{Milton1}. In calculations performed according to Lifshitz
approach and its modifications, cut-off dependent terms were found
to survive in all regularization schemes (see the Appendix in
\cite{Mar}, for example). Only in the special case of the magnetic
dielectric ball satisfying the condition $\varepsilon \mu =1 $,
$\mu$ is a magnetic permeability, Brevik and Kolbenstvedt found
that cut-off terms don't appear in the theory \cite{Brevik1}. The
physical reason for appearance of cut-off dependent terms remained
unclear for a long period. Hoye and Brevik argued in \cite{Hoye}
that the problem has its origin in a continuum description of
dielectric media and can be resolved only in a proper microscopic
model involving minimum separation between molecules.

From the {\it microscopic} viewpoint the Casimir energy of a
dielectric media is usually assumed to arise due to many-body
interatomic interactions of neutral atoms. In the model considered
in the current paper all atoms constituting a dielectric ball are
separated by distances greater than $\lambda$. This is why if one
calculates the total ball energy due to some realistic potential
between atoms of the ball, one always gets the finite energy. The
question is which potential between atoms should be used. The
following line of reasoning is appropriate for dilute connected
dielectrics \cite{Mar}. For the system consisting of a homogeneous
dielectric with the permittivity $\varepsilon(i\omega)$ and a
particle (atom) with the polarizability $\alpha(i\omega)$ located
outside this dielectric it was rigorously proved in \cite{Mar}
that the Casimir energy of this system (the part of it describing
interaction of the atom with a dielectric body in the lowest order
$(\varepsilon(i\omega) - 1)\,\alpha(i\omega)$) coincides with the
sum of pairwise interactions via a dipole-dipole potential
(\ref{tt1}) between the atom and the dielectric body. For the
model of a dielectric ball considered in this paper one can
imagine an atom inside the ball and draw a sphere of radius
$\lambda$ around this atom. Outside this sphere a dilute
dielectric can be described well by $\varepsilon(i\omega)$.
Casimir energy of interaction between an atom and a dielectric
located outside the $\lambda$-sphere is equal in the order
$(\varepsilon(i\omega) - 1)\,\alpha(i\omega)$ to the sum of
dipole-dipole pairwise interactions between the atom and points of
the ball outside the $\lambda$-sphere. To find the total Casimir
energy of atoms constituting the ball in the order
$(\varepsilon(i\omega) - 1)^2$, one has to perform this trick for
every atom of the ball, sum over all atoms and finally multiply
this sum by a factor $1/2$. This procedure is applied to a dilute
dielectric ball in the next section. The interatomic distance
$\lambda$ is a physical cut-off.

Recently the microscopic approach based on two-point correlation
functions and form-factors of different bodies was developed by
Barton \cite{Barton}. The leading terms for short and long
distance contributions to the Casimir energy of different bodies
were derived by Barton by use of two-point correlation functions
in the case of a simple-harmonic oscillator model for dielectric
permittivity. The leading terms for arbitrary model of dielectric
permittivity in the case of a dilute ball can be obtained from our
formula (\ref{tt3}); in the special case of a simple-harmonic
oscillator model for dielectric permittivity they are in agreement
with Barton.

In the next section it is proved that Casimir surface force on a
dielectric ball in the order $(\varepsilon(i\omega) - 1)^2$ is
attractive. For a perfectly conducting spherical shell the force
is repulsive. Brevik and Kolbenstvedt observed \cite{Brevik1} that
energy of a perfectly conducting spherical shell can be derived
from the energy of a magnetic dielectric ball satisfying
$\varepsilon\mu = 1$ in the limit $\mu \to 0$. One may ask a
question. If we first put $\mu = 1$ and after that perform the
limit $\varepsilon \to \infty$, will we obtain the repulsive force
or the attractive contribution will be dominating ? It is
impossible to answer this question at the moment, the theory for
arbitrary $\varepsilon(i\omega)$ still doesn't exist for connected
dielectrics. Microscopic  structure of materials should be an
important element of such a theory, though the complete theory may
probably be developed only at the intersection of macroscopic and
microscopic approaches to the ground state energy.

\section{Casimir energy of dilute ball}

In this section we first calculate the Casimir energy of neutral
atoms constituting a ball of radius $a$ due to a dipole-dipole
pairwise interaction. Then we prove that Casimir surface force is
attractive for arbitrary physically possible dielectric
permittivity $\varepsilon(i\omega) $ of  the ball.

A dipole-dipole interaction  of two neutral atoms with atomic
polarizabilities $\alpha_1(i \omega)$ and $\alpha_2(i \omega)$ is
described by the potential \cite{Dzialoshinskii}
\begin{equation}
U(r) = - \frac{1}{\pi r^2} \int_{0}^{+ \infty} \,
\omega^4 \alpha_1(i \omega) \, \alpha_2 (i \omega)\,  e^{-2\, \omega r }
\Bigl[1+ \frac{2}{\omega r} + \frac{5}{(\omega r)^2 }
+ \frac{6}{(\omega r)^3} + \frac{3}{(\omega r)^4} \Bigr] d\omega , \label{tt1}
\end{equation}
where $r$ is a distance between two atoms. The energy calculation
is illustrated by Fig.$1$. Suppose that an atom (a molecule) with
an atomic polarizability $\alpha(i\omega)$ is located at the point
$B$. One has to integrate interaction of an atom at the point $B$
via a potential (\ref{tt1})  with the atoms separated by distances
greater than interatomic distances $\lambda$ from the point $B$,
integrate over all atom locations $B$ inside the ball and multiply
by a factor $1/2$ to calculate the energy. Assuming homogeneity of
the ball (it results in
$\alpha_1(i\omega)=\alpha_2(i\omega)=\alpha(i\omega)$ and the
condition that the number density of atoms $\rho$ doesn't depend
on the point inside the ball), the Casimir energy is equal to
\begin{multline}
E = \frac{\rho^2}{2} \Biggl(\int_{0}^{a-\lambda} dp\, 4\pi p^2 \int_{\lambda}^{a-p}
dr\, 4\pi r^2 U(r) + \\ \int_{0}^{a-\lambda} dp\, 4\pi p^2
\int_{a-p}^{a+p} dr \, 2\pi r^2 \Bigl(1-\frac{r}{2p}-
\frac{p^2-a^2}{2pr}\Bigr) U(r) + \\
\int_{a-\lambda}^{a} dp \, 4\pi p^2 \int_{\lambda}^{a+p} dr\,
2\pi r^2 \Bigl(1-\frac{r}{2p}- \frac{p^2-a^2}{2pr}\Bigr) U(r)
 \Biggr)  \label{tt2}
\end{multline}

\begin{figure}
\begin{center}
\includegraphics[height=6cm]{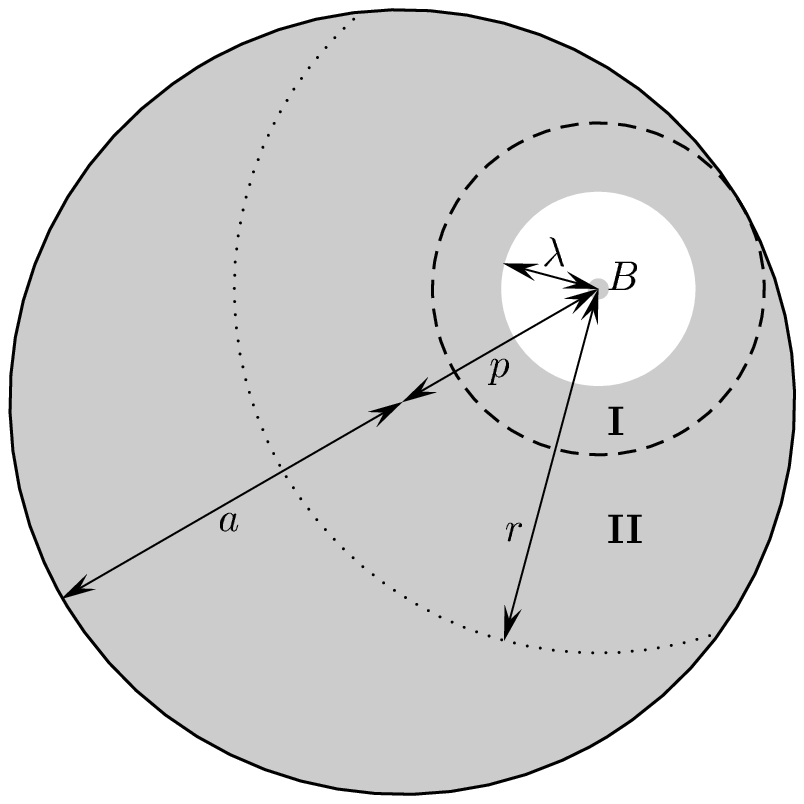} \hfill
\includegraphics[height=6cm]{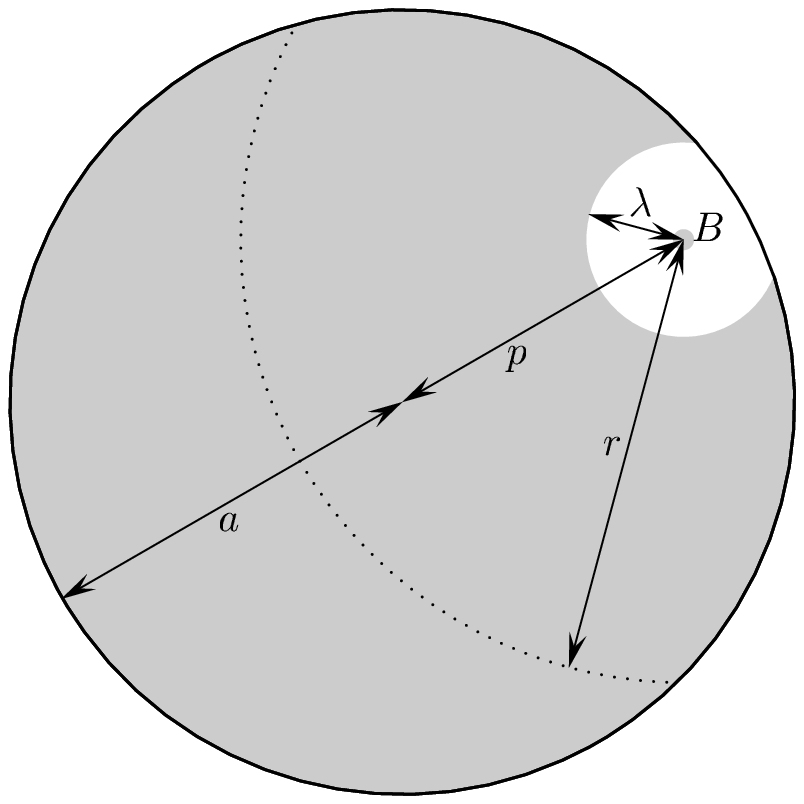} \hfill

\vspace{0.5cm} Fig. 1: Energy calculation for the ball. Shaded
areas $\bf \Large I$ and $\bf \Large II$ separated by a dashed
line on the left picture denote areas of integration in the first
and second integrals in (\ref{tt2}) respectively. A shaded area on
the right picture represents the area of integration in the third
integral in (\ref{tt2}). A part of the sphere of radius $r$ which
is located inside the dielectric ball is denoted by a dotted line.
Its area is equal to $2\pi r^2 (1 - (p^2+r^2-a^2)/(2pr))$.
\end{center}
\end{figure}

Performing calculations, it is straightforward to obtain
\begin{equation}
\begin{split}
E= & - \rho^2 \frac{\pi}{48} \int_{0}^{+\infty} d\omega \alpha^2(i\omega) \,
 \Bigl( \frac{a^3}{\lambda^3} e^{-2\omega\lambda} (128 + 256 \omega\lambda +
 128 \omega^2\lambda^2  + 64 \omega^3\lambda^3 ) - \\
 & -\frac{a^2}{\lambda^2} \bigl(e^{-2 \omega\lambda}(144 + 288\omega\lambda +
 120\omega^2\lambda^2 + 48\omega^3\lambda^3 ) - 96\omega^2\lambda^2
 E_1(2\omega\lambda) \bigr) +  \\ & \bigl(e^{-2\omega\lambda}(41 +
 34\omega\lambda + 14\omega^2\lambda^2 + 4\omega^3\lambda^3 ) +
24 E_1(2\omega\lambda)\bigr) + \\ & \bigl(e^{-4\omega a} (-21 + 12\omega a) -
 E_1(4\omega a) (24 + 96 \omega^2 a^2) \bigr)   \Bigr) ,
\end{split}  \label{tt3}
\end{equation}
where $E_1(x)=\int_{1}^{+\infty} e^{-tx}/t \, dt$.

The formula (\ref{tt3}) answers one question which has longly been
discussed. The question was: is it possible to make the Casimir
energy of connected dielectrics finite (work without divergences)
by selecting the proper model of a frequency dependent dielectric
permittivity $\varepsilon(i\omega)$ (with a quickly decreasing
behaviour when $\omega \to \infty$) in a macroscopic approach
without other assumptions ? {\it The answer is : there is no such
a model}. Only when a finite separation between atoms $\lambda$ is
taken into account, the energy is finite and physical.  When
$\lambda \to 0$, the leading term in (\ref{tt3}) ($V$ is a ball
volume) is
\begin{equation}
- \rho^2 \frac{2 V }{\lambda^3} \int_{0}^{+\infty} d\omega \,
 \alpha^2 (i\omega), \label{tt4}
\end{equation}
so if we wish to take the hypothetical limit $\lambda \to 0$, we
obtain a divergence for every model of $\alpha(i\omega)$. The
leading term (\ref{tt4}) can also be obtained directly from the
van der Waals limit of (\ref{tt1}) for every connected dielectric.

The structure of (\ref{tt3}) clearly shows that $4$ types of terms
contribute to the energy. First and second lines of (\ref{tt3})
represent volume and surface contributions to the energy
respectively. The third line consists of terms which don't depend
on the radius of the ball. The fourth line consists of terms which
don't depend on the atomic separation $\lambda$ and depend only on
the ball radius $a$.

According to general properties of dielectric permittivity on an
imaginary axis, it is possible to write the leading contribution
from the fourth line of (\ref{tt3}) as (it comes from frequencies
$\omega \ll \omega_0$, $\omega_0$ is a characteristic absorption
frequency of materials, $\omega_0 a \gg 1$, so it is possible to
use the static polarizability $\alpha(0)$ in the leading
approximation to the fourth line of (\ref{tt3}) )
\begin{multline}
- \rho^2 \alpha^2(0) \frac{\pi}{48} \int_{0}^{+\infty} d\omega  \,
  \bigl(e^{-4\omega a} (-21 + 12\omega a) -
 E_1(4\omega a) (24 + 96 \omega^2 a^2) \bigr) = \\
= \rho^2 \alpha^2(0) \frac{23}{96} \frac{\pi}{a}  \label{tt5}
\end{multline}
A review and references to different approaches which had been
used to derive (\ref{tt5}) are given in the Appendix of
\cite{Mar}.

However, the term  (\ref{tt5})  has a small influence on physics
because physics is mainly governed by (\ref{tt4}) and next to
leading terms in (\ref{tt3}).  The reason is quite simple -  the
term (\ref{tt5}) is much less in magnitude than the terms in the
first, second and  third  lines of (\ref{tt3}). For example, ratio
of terms in the first line of (\ref{tt3}) to (\ref{tt5}) is $\sim
(\omega_0 a) (a/\lambda)^3 \gg 1$. Usually all terms but
(\ref{tt5}) which appeared in different renormalization schemes
were simply discarded or considered in macroscopic approaches as
divergent contributions which  add up to macroscopic quantities of
the ball (e.g. volume, surface  energies) during the
renormalization procedure, as it often happens in field theory. So
only the term (\ref{tt5}) was usually considered as the Casimir
energy term. The coefficient $a_2$ in the asymptotic expansion of
the heat kernel for a nondispersive dielectric ball in the order
$(\varepsilon - 1)^2$ vanishes \cite{Vas}, this explains the fact
that no logarithmic terms appeared in macroscopic calculational
schemes for  the ball energy, so it was possible to separate the
term (\ref{tt5}) uniquely  from the divergent part of the energy.
Theoreticians got used to believe that the Casimir energy should
depend only on geometry of the bodies, not on their internal
structure,  the term (\ref{tt5}) depended only on the ball radius
$a$, so it seemed to be a proper Casimir energy term. However,
Casimir surface force resulting from (\ref{tt5}) is repulsive.
This consequence seemed unusual but possible. Only the equivalence
of the Casimir energy and a dipole-dipole interaction for
homogeneous dielectrics in the order $(\varepsilon(i\omega) -
1)^2$, proved in \cite{Mar} on the basis of Lifshitz theory, makes
obvious the fact that Casimir surface force should be attractive.
The potential (\ref{tt1}) of a dipole-dipole interaction is
attractive for all distances, and one can not obtain repulsive
Casimir forces for dilute dielectrics in principle because one
starts from the attractive potential. Even from this argument it is
clear that the term (\ref{tt5}) is not enough to obtain physically
meaningful results. Terms depending on the interatomic distance
$\lambda$ are greater in magnitude than (\ref{tt5})  and must be
taken into account in the derivation of Casimir surface force.

After these remarks it is natural to prove  that Casimir surface
force on a dilute dielectric ball is attractive. It is convenient
to define $N\equiv a/\lambda, p\equiv\omega\lambda$. Then the
formula (\ref{tt3}) can be rewritten in a general form
\begin{equation}
E = - \frac{\rho^2}{\lambda} \int_{0}^{+\infty} dp \, \alpha^2\Bigl(i
\frac{p}{\lambda}\Bigr) f(N,p)    \label{tt6}
\end{equation}
The function $f(N,p)>0$ for $N>1, p>0$. The ball expands or
collapses homogeneously, so
\begin{equation}
N=const. \label{tt7}
\end{equation}
Conservation of atoms inside
the ball imposes the condition
\begin{equation}
\rho \, \frac{4\pi a^3}{3} = const. \label{tt8}
\end{equation}
It is convenient to use  Kramers-Kronig relations in the form
\begin{equation}
\alpha(i\omega) = \int_{0}^{+\infty} dx\, \frac{x g(x)}{x^2+\omega^2}, \label{tt9}
\end{equation}
where the condition $g(x)>0$ always holds \cite{Landau}.
Using (\ref{tt6}), (\ref{tt7}), (\ref{tt8}), (\ref{tt9}), Casimir
force on a unit surface is equal to
\begin{multline}
F = -\frac{1}{4\pi a^2} \frac{\partial E}{\partial a} = \\ =  -
\frac{\rho^2}{4\pi a^3}\int_{0}^{+\infty} d\omega \int_{0}^{+\infty}
dx \, \frac{x (7x^2 + 3 \omega^2) g(x)}{(x^2+\omega^2)^2}\,
\alpha(i\omega) f(N,\omega \lambda)    < 0 , \label{tt10}
\end{multline}
$F<0$ because all functions inside  integrals are positive. Casimir
surface force is attractive for every model of atomic
polarizability consistent with general causal requirements.

It may be worth imagine a model of a dilute dielectric ball which
wouldn't exist at all without a dipole-dipole interaction. Without
a dipole-dipole interaction  atoms would be free at large
separations, they would interact only during the collisions due to
short range interatomic forces. Energy of a dipole-dipole
interaction of such a system is just the energy which holds atoms
of the ball together. When the energy due to a dipole-dipole
atomic interaction is equal to the repulsive energy  due to short
range interatomic forces, the system is in equilibrium and stable.

\section{Summary}
The Casimir energy of a dilute homogeneous nonmagnetic dielectric
ball at zero temperature is derived for the first time for an arbitrary physically
possible frequency dispersion of a dielectric permittivity in the order
$(\varepsilon(i\omega) - 1)^2$ by summing up dipole-dipole
interactions between atoms constituting the ball. All calculations
are performed without divergences because average interatomic distance
$\lambda$ is a {\it physical} cut-off.
Casimir surface force is proved to be attractive.

\section{Acknowledgments}
Author acknowledges valuable discussions and correspondence with
Gabriel Barton, Michael Bordag, Iver Brevik, Yuri Novozhilov and
Dmitri Vassilevich.


\begin{thebibliography}{99}

\bibitem{Mar} V.N.Marachevsky , Casimir energy and dilute dielectric
 ball , \\ hep-th/0010214, to appear in Physica Scripta.

\bibitem{Lif1} E.M.Lifshitz, Zh.Eksp.Theor.Fiz.{\bf 29}, 94 (1955).

\bibitem{Lifshitz}   E.M.Lifshitz and  L.P.Pitaevskii, {\it Statistical
Physics, Part 2} (Course of Theoretical Physics, vol. IX), (Moscow,
Nauka, 1978), Chapter 8.

\bibitem{M1}
K.A.Milton and  Y.J.Ng,  Phys.Rev.E  {\bf 57}, 5504 (1998).

\bibitem{tech}
I.Brevik and V.Marachevsky, Phys.Rev.D {\bf 60}, 085006 (1999).

I.Brevik, V.N.Marachevsky  and  K.A.Milton,
 \\ Phys.Rev.Lett.{\bf 82}, 3948 (1999), hep-th/9810062.

G.Barton, J.Phys. A {\bf 32} , 525 (1999).

\bibitem{Boyer}
T.H.Boyer, Phys.Rev. {\bf 174}, 1764 (1968).

\bibitem{Milton2}
K.A.Milton, L.L.DeRaad,Jr. and  J.Schwinger,
Ann.Phys.(N.Y.) {\bf 115}, 388 (1978).

\bibitem{Milton1}
K.A.Milton, Ann.Phys.(N.Y) {\bf 127}, 49 (1980).

\bibitem{different}
Some articles presenting various aspects and techniques in Casimir effect:

Yu.S.Barash and  V.L.Ginzburg, Usp.Fiz.Nauk {\bf 116}, 5 (1975).

K.A.Milton, L.L.DeRaad,Jr. and  J.Schwinger,
Ann.Phys.(N.Y.) {\bf 115}, 1 (1978).

M.Bordag, K.Kirsten and D.V.Vassilevich, J.Phys.A {\bf 31}, 2381
(1998), hep-th/9709084.

V.V.Nesterenko and I.G.Pirozhenko, Phys.Rev.D {\bf 60}, 125007 (1999), \\ hep-th/9907192.

I.Klich, Phys.Rev.D {\bf 61}, 025004 (2000), hep-th/9908101.

J.S.H{\o}ye, I.Brevik and J.B.Aarseth, The Casimir problem of Spherical
Dielectrics: Quantum Statistical and Field Theoretical Approaches,
quant-ph/0008088.

H.Falomir, K.R\'{e}bora and M.Loewe, Phys.Rev.D {\bf 63}, 025015
(2001), \\hep-th/0008251. 

H.Falomir, K.Kirsten and K.R\'{e}bora, Divergencies in the Casimir
energy for a medium with realistic ultraviolet behaviour,
hep-th/0103050.

E.M.Santangelo, Evaluation of Casimir energies through spectral
functions, hep-th/0104025. 
   

\bibitem{Brevik1} I.Brevik and H.Kolbenstvedt,
Ann.Phys.(N.Y.) {\bf 143}, 179 (1982);  {\bf 149}, 237 (1983).

\bibitem{Hoye} J.S.H{\o}ye and I.Brevik,
J.Stat.Phys. {\bf 100}, 223 (2000), quant-ph/9903086.

\bibitem{Barton}
G.Barton, Perturbative Casimir energies of dispersive spheres, cubes,
and cylinders, preprint, November 2000.


\bibitem{Dzialoshinskii} V.B.Berestetskii, E.M.Lifshitz and
L.P.Pitaevskii ,   {\it Quantum Electrodynamics}
(Course of Theoretical Physics, vol. IV ), third ed., (Moscow, Nauka,
1989), Chapter 9, eq.(85.17).

\bibitem{Vas}
M.Bordag, K.Kirsten and D.V.Vassilevich, Phys.Rev.D {\bf 59}, 085011
(1999),  hep-th/9811015.


\bibitem{Landau} L.D.Landau and E.M.Lifshitz, {\it Electrodynamics of
 continuous media} (Course of Theoretical Physics, vol. VIII), second
 ed., (Moscow, Nauka, 1982), Chapter 9, eq.(82.15).



\end{thebibliography}
\end{document}